\documentstyle[aps,epsfig,12pt]{revtex}
\renewcommand{\theequation}{\arabic{section}.\arabic{equation}}

\begin{document}
\title{$B(H)$ Constitutive Relations Near $H_{c1}$ in Disordered
Superconductors}
\author{Raphael A. Lehrer and David R. Nelson} 
\address{Lyman Laboratory of Physics, Harvard University, Cambridge,
Massachusetts 02138}
\date{August 8, 1999}
\maketitle

\begin{abstract}
We provide a self-contained account of the $B$ vs.\ $H$ constitutive
relation near $H_{c1}$ in Type II superconductors with various types
of quenched random disorder.  The traditional Abrikosov result $B \sim
[ \ln (H-H_{c1})]^{-2}$, valid in the absence of disorder and thermal
fluctuations, changes significantly in the presence of disorder.
Moreover, the constitutive relations will depend strongly on the type
of disorder.  In the presence of point disorder, $B \sim (H -
H_{c1})^{3/2}$ in three-dimensional (thick) superconductors, as shown
by Nattermann and Lipowsky.  In two-dimensional (thin film)
superconductors with point disorder, $B \sim (H - H_{c1})$.  In the
presence of parallel columnar disorder, we find that $B \sim \exp
[-C_3 / (H - H_{c1})]$ in three dimensions, while $B \sim \exp [-C_2 /
(H - H_{c1})^{1/2}]$ in two dimensions.  In the presence of nearly
isotropically splayed disorder, we find that $B \sim (H -
H_{c1})^{3/2}$ in both two and three dimensions.
\end{abstract}

\section{Introduction}
\label{sec-introsplay}

The physics of vortex lines in high-temperature superconductors has
attracted much experimental and theoretical interest~\cite{Blatter}.
The competition between interactions, pinning, and thermal
fluctuations gives rise to a wide range of novel phenomena that are
both interesting in their own right and technologically important.
Here, we focus primarily on the effects of disorder on vortex
behavior, phenomena which are also important for low-temperature Type
II superconductors.  In a disorder-free sample, vortex lines always
flow in response to a current, leading to a resistance even at
arbitrarily small currents.  By contrast, defects in superconductors
attract vortex lines.  Just as a few nails can hold a carpet in place,
an entire vortex line system can be held in place by a few defects,
leading to a large critical current---provided the vortices are in a
solid phase, where they are held in place relative to each other by
their mutual repulsion, which gives rise to a shear modulus.  If the
temperature is raised so that the vortices are in a liquid state, then
vortices held in place by the disorder are pinned, but the other
vortices can flow around them, leading again to zero critical current.

In the case of the high-temperature superconductors, disorder is
naturally present in the form of oxygen vacancies.  The quantity of
such vacancies can be altered by changing the doping of the crystal,
\textit{i.e.}, by introducing more or less oxygen during the growth
process.  Such changes can have a strong impact on the pinning of the
vortices, and hence on the critical current.  Substitutional defects
can play a similar role in low-temperature superconductors.

Artificial defects added to superconductors are particularly effective
in increasing the critical current.  Of special note are columnar
defects, which are generated by bombarding the sample with heavy ions
that produce damage tracks in their
wake~\cite{Hardy,Civale,Konczykowski,Gerhauser,Budhiani}.  These
tracks pin vortices strongly because their width ($\sim 60 \AA$) is
comparable to the vortex core size ($\sim 20 \AA$).  When the columnar
pins are aligned with the direction of the magnetic field (and hence
with the vortex lines) a large increase in the critical current is
observed.

However, an even greater increase in the critical current is observed
when the columns are splayed, \textit{i.e.}, they are not all oriented
in the same direction~\cite{Civale2,Kruzin,Hardy2}.  The superior
pinning properties of columnar defects with controlled splay was
predicted theoretically based on expected reduction in a
variable-range hopping vortex transport mechanism and enhanced vortex
entanglement due to splay~\cite{Hwasplay,LeDoussal+N}.  Despite the
technological importance, much remains to be done in order to
understand the behavior of vortex lines in the presence of splayed
columnar disorder.

Even at high temperatures, when the analysis is expected to be most
tractable, standard approaches seem to break down and give nonsensical
results.  For example, the boson
mapping~\cite{N88,N+Vinokur,N+LeDoussal} works quite well in
describing many of the properties of vortex lines in the presence of
point or unsplayed columnar disorder.  This approach utilizes a formal
correspondence between vortex trajectories and the world lines of
fictitious quantum-mechanical bosons in two dimensions.  In this
analogy, the temperature $T$ plays the role of Planck's constant
$\hbar$, the bending energy (or line tension) $g$ plays the role of
the boson mass $m$, and the length of the sample $L$ corresponds to
$\beta \hbar$ for the bosons.  The analogy works best in the $L
\rightarrow \infty$ limit, which corresponds to the $T \rightarrow 0$
limit for the bosons.  In this limit, in the absence of disorder and
interactions, the bosons should form a condensate.  In the presence of
disorder or interactions, some ``bosons'' are kicked out of the
condensate, resulting in a condensate density $n_0$ which is less than
the total density $n$.  For vortex lines, the ``condensate density''
is a measure of the degree of
entanglement~\cite{N88,N+Vinokur,Tauber+N}.  This condensate density
can be calculated for the various types of disorder discussed above.
It is well-behaved for point and unsplayed columnar disorder, but in
the presence of even weak splayed columnar disorder, $n_0$
diverges~\cite{Tauber+N}, suggesting that the boson mapping is flawed
in the presence of splayed columnar disorder.  Indeed, T\"{a}uber and
Nelson~\cite{Tauber+N} conclude that the super-diffusive wandering of
the flux lines causes the mapping onto non-relativistic bosons to
break down.  Unfortunately, this mapping---which has provided many
insights for the cases of no disorder, point disorder, and unsplayed
columnar defects---seems to be less suitable for understanding the
behavior of vortex lines in the presence of splayed columnar disorder.

Here, we focus on the behavior of vortex lines near $H_{c1}$, where
the lines are dilute.  In particular, we predict the $B(H)$
constitutive relation for vortex lines in the presence of the various
types of disorder discussed above.  We begin by reviewing the
traditional Abrikosov result, expected to hold in the absence of
disorder and thermal fluctuations.  Each vortex that enters the sample
will gain a free energy proportional to $(H-H_{c1})$ per unit length.
However, there is an energy cost due to repulsive interactions between
any two vortices proportional to 
$\frac{1}{\sqrt{r}} \exp (-r/{\lambda})$ for $r \gg \lambda$, where
$r$ is the distance between the vortices and $\lambda$ is the London
penetration depth.  In the dilute limit, the interactions with nearest
neighbors will dominate over interactions with more distant neighbors,
and the free energy density is given by
\begin{equation}
f = - c_1 (H-H_{c1}) n + c_2 n^{5/4} e^{-c_3 / \sqrt{n}},
\end{equation}
where $n$ is the density of vortices, and $c_1$, $c_2$, and $c_3$ are
constants that can be determined in terms of the vortex
parameters~\cite{Tinkham}.  (We leave them general here to better
elucidate the structure of the argument.)  Upon minimizing $f$ with
respect to $n$, we obtain
\begin{equation}
n = \left\{ \frac{c_3}{\ln \left[ \frac{c_2 c_3}{2 c_1 n^{1/4}}
\frac{1}{(H-H_{c1})} \right]} \right\}^2.
\end{equation}
The dominant behavior may be obtained by substituting $n \propto
c_3^2$ on the right hand side, so that the magnetic field (given by $B
= n \phi_0$, where $\phi_0$ is the flux quantum) varies inversely as
the square of the logarithm of $H-H_{c1}$.  Plugging in the relevant
parameters $c_1$, $c_2$, and $c_3$ for a triangular lattice, one
finds~\cite{Parks2,Tinkham}
\begin{equation}
n = \frac{2\phi_0}{\sqrt{3} \lambda^2} \left\{ \frac{1}{\ln \left[
\frac{3 \phi_0}{4\pi \lambda^2} \frac{1}{(H-H_{c1})} \right]} \right\}^2.
\label{eq:abrikosov}
\end{equation}

\begin{figure}[tp]
\begin{center}\leavevmode
\includegraphics[width=0.4\linewidth]{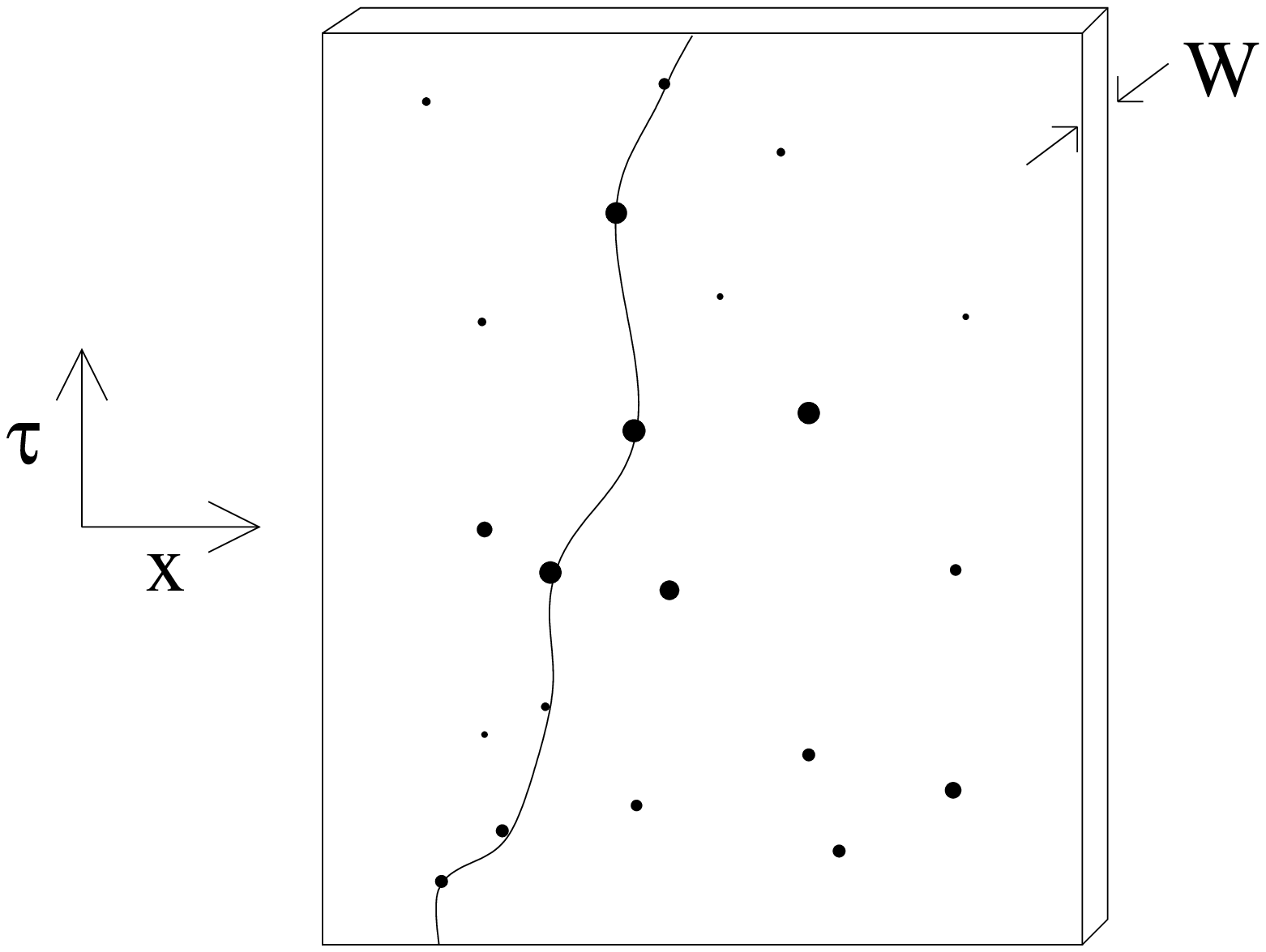}
\caption{A vortex line in a 1+1 dimensional sample of width $W$ in the
presence of point disorder.}
\label{fig:point2d}
\end{center}
\end{figure}

\begin{figure}[tp]
\begin{center}\leavevmode
\includegraphics[width=0.4\linewidth]{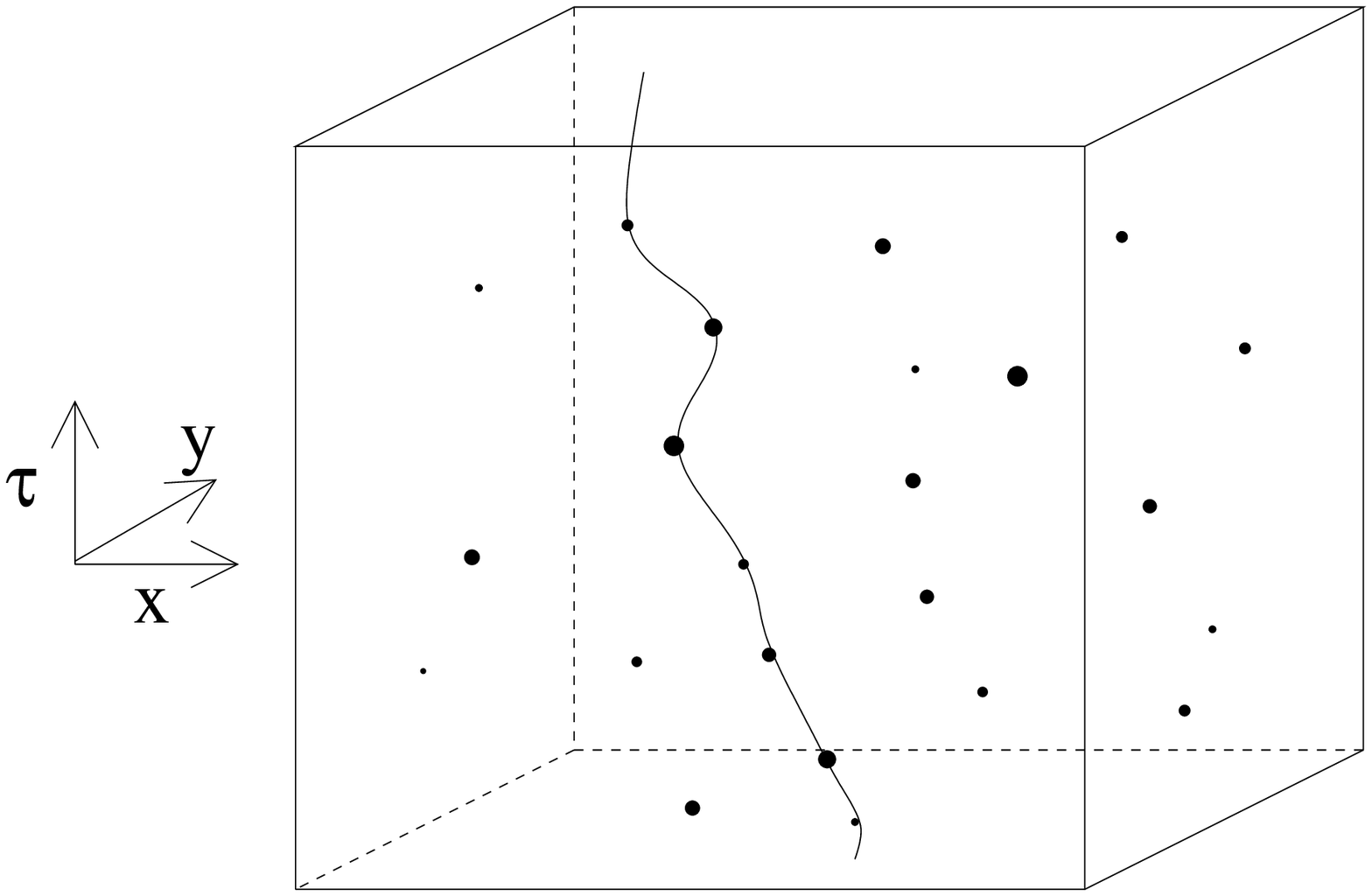}
\caption{A vortex line in a 2+1 dimensional sample in the presence of
point disorder.}
\label{fig:point3d}
\end{center}
\end{figure}

\begin{figure}[tp]
\begin{center}\leavevmode
\includegraphics[width=0.4\linewidth]{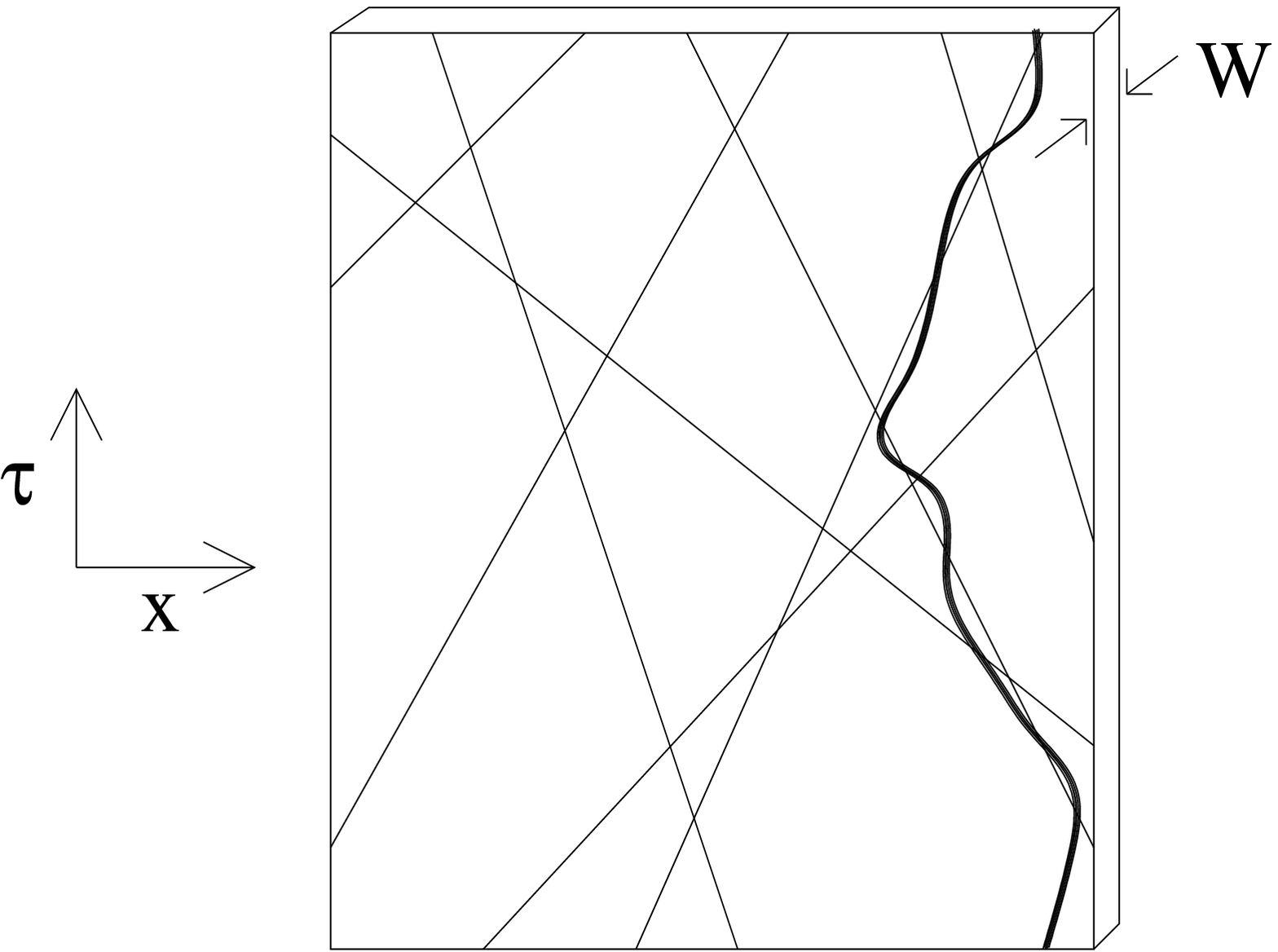}
\caption{A vortex line in a 1+1 dimensional sample of width $W$ in the
presence of splayed columnar disorder.}
\label{fig:splay2d}
\end{center}
\end{figure}

\begin{figure}[tp]
\begin{center}\leavevmode
\includegraphics[width=0.4\linewidth]{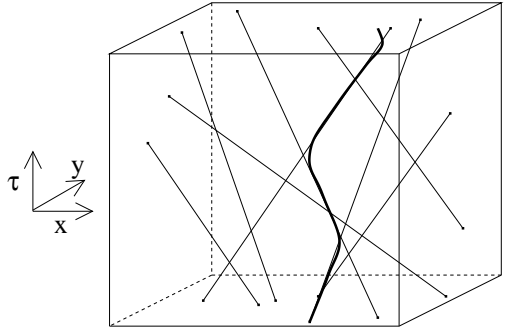}
\caption{A vortex line in a 2+1 dimensional sample in the presence of
splayed columnar disorder.}
\label{fig:splay3d}
\end{center}
\end{figure}

\begin{figure}[tp]
\begin{center}\leavevmode
\includegraphics[width=0.4\linewidth]{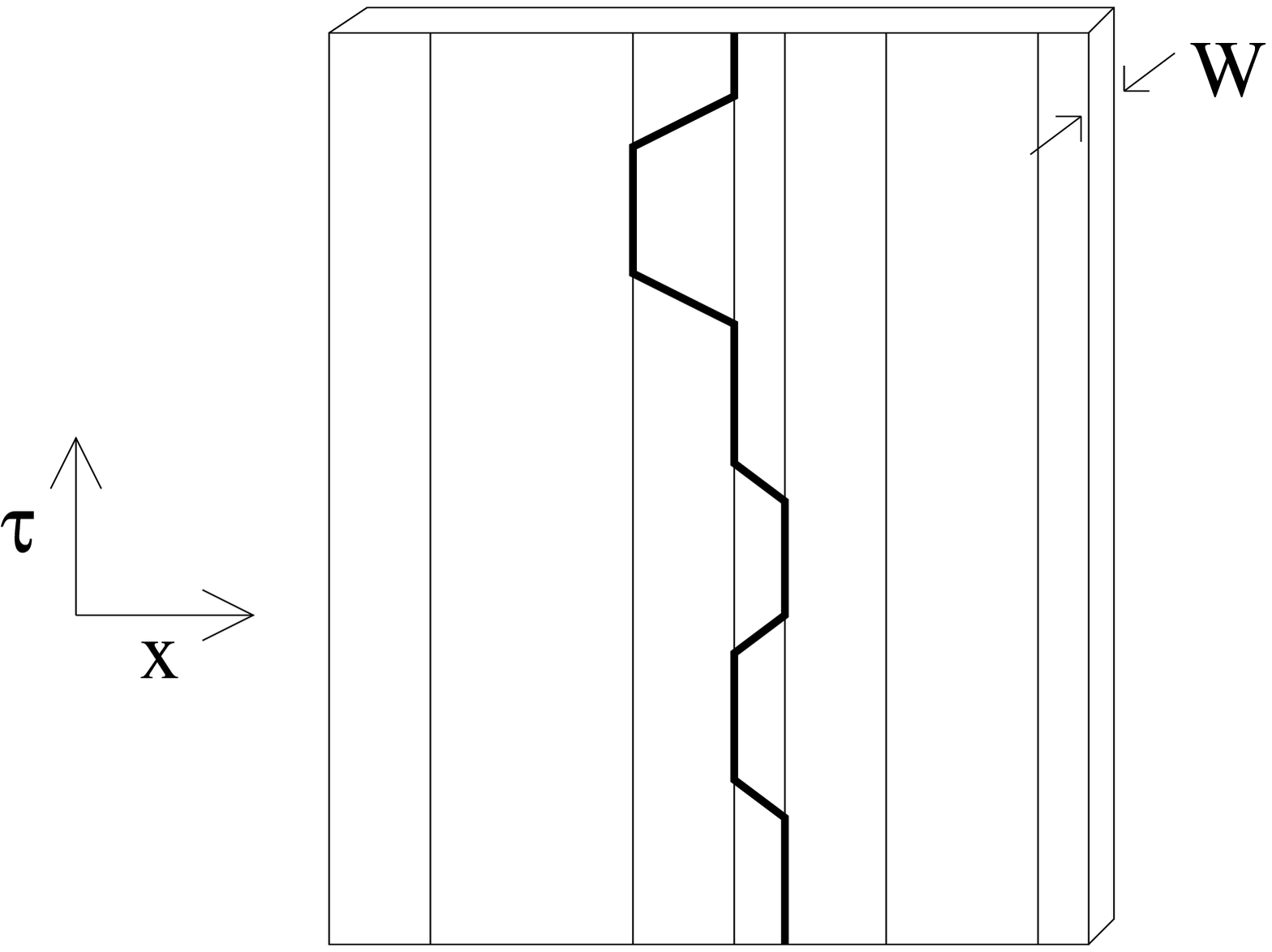}
\caption{A vortex line in a 1+1 dimensional sample of width $W$ in the
presence of parallel columnar disorder.}
\label{fig:column2d}
\end{center}
\end{figure}

\begin{figure}[tp]
\begin{center}\leavevmode
\includegraphics[width=0.4\linewidth]{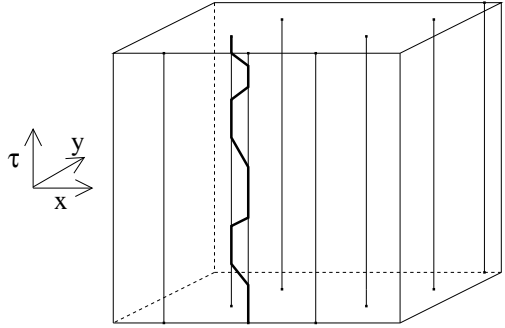}
\caption{A vortex line in a 2+1 dimensional sample in the presence of
parallel columnar disorder.}
\label{fig:column3d}
\end{center}
\end{figure}

In the presence of disorder, Eq.\ (\ref{eq:abrikosov}) will be
modified.  With point disorder, we obtain $B \sim (H - H_{c1})$ (as
calculated~\cite{expt,Kardar} and measured by Bolle \textit{et
al}.~\cite{expt}) in 1+1 dimensional samples, where the vortices only
have one direction transverse to the magnetic field in which they can
wander (see Fig.~\ref{fig:point2d}).  In 2+1 dimensional samples
(Fig.~\ref{fig:point3d}), we obtain $B \sim (H - H_{c1})^{3/2}$, in
agreement with calculations done by Nattermann and
Lipowsky~\cite{Nattermann}.  In the presence of splayed columnar
defects, we find that $B \sim (H - H_{c1})^{3/2}$ both for 1+1
dimensional (Fig.~\ref{fig:splay2d}) and 2+1 dimensional
(Fig.~\ref{fig:splay3d}) samples.  For columnar disorder with unbound
disorder strength, Larkin and Vinokur~\cite{Larkin+Vinokur} argue that
$B \sim e^{C (H - H_{c1})}$ in 2+1 dimensions.  However, we show that
for the more physical case of bounded disorder, $B \sim \exp [-C_3 /
(H - H_{c1})]$ in 2+1 dimensions (Fig.~\ref{fig:column2d}), while $B
\sim \exp [-C_2 / (H - H_{c1})^{1/2}]$ in 1+1 dimensions
(Fig.~\ref{fig:column3d}).  In addition to these relations, we also
estimate the prefactors (up to factors of order unity) in terms of
physical constants such as the temperature, the disorder strength, and
the superconducting coherence length.  These prefactors are important
for comparisons with experiment.
\pagebreak

A recent torsional oscillator experiment carried out by Bolle
\textit{et al}.~\cite{expt} confirms that $B \sim (H - H_{c1})$ for
vortex lines in two dimensions in the presence of point disorder.  The
experiment was done by attaching a thin sheet of superconducting
2H-NbSe$_2$ to a high-$Q$ micro-electromechanical device, and
measuring the resonance frequency of torsional oscillations of the
sample.  Because a sample with a magnetic moment ${\bf M}$ in a field
${\bf H}$ exerts an additional torque ${\bf \tau} = {\bf M} \times
{\bf H}$ on the oscillator, the magnetic field which penetrates and
becomes trapped in the sample can be probed by finding shifts in the
resonance frequency as a function of applied field $H$.  Small jumps
in the frequency as a function of applied field were
observed~\cite{expt}, which were attributed to individual vortices
entering the sample.  By counting the number of jumps as a function of
$H$, these authors obtained $B \sim (H - H_{c1})$.  In the presence of
parallel or splayed columnar disorder, the $B$ vs.\ $H$ curve could be
measured by a very similar experiment, where the sample was first
irradiated isotropically by heavy ions to produce the required
disorder.  In three dimensions, a similar experiment could be done
using a long thin needle-shaped sample to avoid demagnetizing effects.
For point disorder, this experiment was done in the 1960's with
ambiguous results~\cite{Parks}.  In particular, work by Finnemore
\textit{et al}.~\cite{Finnemore} on Niobium samples seems to indicate
$B \sim (H - H_{c1})^x$, with $x > 1$.  However, the decades-old data
is too rough to allow quantitative comparison with the prediction for
point disorder $B \sim (H - H_{c1})^{3/2}$.

We approach these problems differently for the different types of
disorder.  Parallel columnar disorder localizes vortices into finite
transverse regions of the sample in both 1+1 and 2+1 dimensions.
Therefore, at low densities, the vortices can easily avoid paying a
large cost associated with their mutual repulsion if they are located
in different areas of the sample.  In the language of the boson
mapping described above, we can approximate the intervortex
interactions as merely restricting the occupancy of any given
localized state to precisely one boson~\cite{N+Vinokur,Hatano+N}.  The
form of the $B(H)$ constitutive relation is then determined by the low
energy tails of the density of states.  This idea is explored further
in Sec.~\ref{sec-columns}.

By contrast, as discussed in Sec.~\ref{sec-burgers}, in the presence
of point disorder or splayed columnar disorder, noninteracting
vortices will be in extended states.  As such, repeated collisions
between vortices play a crucial role in determining the $B(H)$
constitutive relation, as the intervortex interactions attempt to
localize each vortex into a cage surrounded by their neighbors.  The
energy lost due to restricted vortex wandering will determine $B(H)$.
We investigate this case more fully beginning in
Sec.~\ref{sec-burgers} with the problem of a single flux line
superimposed on a background of disorder.  The partition function
describing a flux line can be mapped onto the noisy Burger's, or KPZ,
equation describing the fluctuations of an elastic interface in the
presence of a random space and time dependent potential influencing
the interface's progress~\cite{HH+Zhang}.  This problem has been
studied extensively for random potentials uncorrelated in space and
time (appropriate to point disorder for the vortex
lines)~\cite{FNS,KPZ,HH+Zhang}, but more recently has also been
investigated for correlated potentials~\cite{MHKZ,HH,FTJ}.  Provided
we restrict ourselves to the case where the splayed columnar defects
are nearly isotropically oriented, we can use previous results for the
wandering exponent $\zeta$, which describes how far the vortex line
wanders transverse to the magnetic field in a distance $l$ in the
field direction by the formula $\left\{ \overline{[{\bf r}(l) - {\bf
r}(0)]^2} \right\}^{1/2}
\sim l^{\zeta}$, where ${\bf r} (z)$ labels the transverse coordinates
of the flux line.  Columnar defects with an approximately isotropic
distribution of splay can be created by using neutrons or protons to
trigger fission of, \textit{e.g.}, bismuth nuclei in
BSCCO~\cite{Krusin2}.

In Sec.~\ref{sec-collisions}, we develop an approximate theory for a
dilute set of vortex lines in the presence of point and splayed
columnar disorder.  We first derive the scaling of $B$ vs.\ $H$, and
then focus on the prefactor of this relation.  In some
experiments~\cite{Civale2,Kruzin,Hardy2}, the columns do not traverse
the entire sample, so we also consider the effect that finite column
size will have on our predictions.  Specifically, we expect a
crossover from the point disorder behavior at extremely weak
$H-H_{c1}$ to the splayed columnar behavior at somewhat larger
$H-H_{c1}$.

\section{Single vortex line}
\label{sec-burgers}
\setcounter{equation}{0}

\subsection{Model}

We begin with a model free energy for a single vortex line in a $d$
dimensional sample of thickness $L$.  We label the direction of the
magnetic field by $\tau$, and the transverse position of the line at
$\tau$ by a $d-1$ dimensional vector ${\bf r}(\tau)$.  The free energy
then reads~\cite{Blatter,N+Vinokur}:
\begin{equation}
F[{\bf r}(\tau)] = \frac{1}{2} g \int_0^L \left( \frac{d {\bf r}}{d\tau}
\right)^2 d\tau + \int_0^L V[{\bf r}(\tau), \tau] d\tau,
\label{eq:dprm}
\end{equation}
where $g$ is the line tension.  The pinning potential $V[{\bf r},
\tau]$ arises due to the interaction with, say, point disorder or
splayed columnar defects.  Its mean value merely affects the average
field $H_{c1}$ at which vortex lines will penetrate the sample, so we
subtract it out, and assume that $\overline{V[{\bf r}, \tau]} = 0$.
We further assume that the noise is Gaussian with a correlator
\begin{equation}
\Delta ({\bf r} - {\bf r'}, \tau - \tau') = \overline{V[{\bf r}, \tau]
V[{\bf r'}, \tau']}.
\end{equation}
We focus on the case of ``nearly isotropic'' splay.  This is the
defect correlator for a set of randomly tilted columnar pins, each
described by a trajectory ${\bf r}(\tau) = {\bf R} + {\bf v} \tau$
with a Gaussian distribution of the tilts ${\bf v}$, $P[{\bf v}]
\propto e^{-v^2/2v_D^2}$ in the limit $v_D \rightarrow
\infty$~\cite{Hwasplay}.  For nearly isotropic splay, the Fourier
transform of the correlator, defined by
\begin{equation}
\Delta ({\bf k}, \omega) = \int d^{d-1} {\bf r} \int_0^L d \tau
\Delta ({\bf r}, \tau) e^{-i({\bf k} \cdot {\bf r} - \omega \tau)},
\end{equation}
is given by $\Delta/k$~\cite{Hwasplay,Tauber+N}, which differs from
the truly isotropic limit $\Delta /(k^2 + \omega^2)^{1/2}$.  However,
using the correlator for nearly isotropic splay simplifies the
calculations significantly.  In the language of the noisy Burger's (or
KPZ) equation, truly isotropic splay has both spatial and temporal
correlations, while nearly isotropic splay has only spatial
correlations.  By focusing on the nearly isotropic splay, we can
therefore rely on work done on the KPZ equation with spatially
correlated disorder and avoid the more complicated case of temporal
disorder.  Moreover, we believe that this difference should not affect
the physical implications at large length scales
significantly~\cite{nearlyisotropic}.

The partition function associated with the free energy of Eq.\
(\ref{eq:dprm}),
\begin{equation}
\mathcal{Z} ({\bf r}, \tau) = \int \mathcal{D} {\bf r}' (\tau ') 
\mathit{e}^{-F[{\bf r}' (\tau ')]/T},
\end{equation}
\textit{i.e.}, the path integral of the Boltzmann factor $e^{- F[{\bf
r}' (\tau ')]/T}$ over all vortex trajectories ${\bf r}' (\tau ')$
running from $\tau ' = 0$ to position ${\bf r}(\tau)$ at $\tau ' =
\tau$, obeys the differential equation
\begin{equation}
T \frac{\partial \mathcal{Z} ({\bf r}, \tau)}{\partial \tau} = \left[
\frac{T^2}{2g} \nabla^2 + V ({\bf r}, \tau) \right] \mathcal{Z}({\bf
r}, \tau).
\label{eq:transfer}
\end{equation}
Eq.\ (\ref{eq:transfer}) can be further transformed by means of the
Cole-Hopf transformation (similar to the WKB transformation in quantum
mechanics)
\begin{equation}
\mathcal{Z}({\bf r}, \tau) = \exp \left[ \mathrm{\Phi} ({\bf r}, \tau)
\right]
\end{equation}
into
\begin{equation}
T \frac{\partial \Phi}{\partial \tau} = \frac{T^2}{2g} \nabla^2 \Phi +
\frac{T^2}{2g} (\nabla \Phi)^2 + V({\bf r}, \tau).
\label{eq:burgervl}
\end{equation}
This equation, known as the KPZ~\cite{KPZ}, or noisy Burger's
equation~\cite{FNS}, has been studied in the case of
\textit{uncorrelated} (point-like) disorder in a great variety of
contexts.  It was studied by Forster, Nelson, and Stephen~\cite{FNS}
as a model for the velocity fluctuations in a randomly stirred,
$(d-1)$-dimensional turbulent fluid.  Later, it reappeared as a model
for surface roughening proposed by Kardar, Parisi, and
Zhang~\cite{KPZ}.  In this context, it has inspired a great deal of
work, summarized in a review article by Halpin-Healy and
Zhang~\cite{HH+Zhang}.

There have also been investigations of the properties of this equation
with correlated disorder $\Delta ({\bf k}) = \Delta / k^{2\rho}$ in
the context of surface roughening.  However, to date, there has not
been any reliable characterization of the exponents for the entire $d$
vs.\ $\rho$ space.  Medina, \textit{et al}.~\cite{MHKZ} have found
results that seem to be accurate in two dimensions~\cite{dimensions}.
Halpin-Healy~\cite{HH} has obtained results that are accurate in two
dimensions, and may describe the behavior for sufficiently small
$\rho$ in higher dimensions as well.  Recently, Frey, \textit{et
al}.~\cite{FTJ} have found exact exponent solutions not only in two
dimensions but also for sufficiently large powers of $\rho$ in higher
dimensions.  (It is by reference to these exact results that we judge
the accuracy of Medina, \textit{et al}. and Halpin-Healy's results.)
Although they are not able to obtain exact results for all $d$ and
$\rho$, Frey, \textit{et al}. make a plausible conjecture for the
behavior at smaller values of $\rho$ that is in agreement with
Halpin-Healy's short range results and with numerical
simulations~\cite{Li}.

Our purpose here is to adapt this work to vortex lines in the presence
of splayed columnar disorder, and explain some of the consequences for
experiments.  The application of the results for point disorder to
experiments~\cite{expt} is also reviewed.  In Sec.~\ref{sec-rgsplay},
we set the stage for a more detailed discussion with a simple,
self-contained renormalization group treatment that gives results in
agreement with Frey, \textit{et al}.'s exact results in its range of
applicability.  In Sec.~\ref{sec-white}, we discuss the limitations of
this approach.

\subsection{Renormalization group}
\label{sec-rgsplay}

\subsubsection{Scaling}
\label{sec-scaling}

To understand the vortex wandering at long length scales, we use the
renormalization group method of Ref.~\cite{FNS} to integrate out the
short-distance behavior.  Since, under renormalization, the
coefficients of the $\nabla^2 \Phi$ term and the $(\nabla \Phi)^2$
will no longer remain identical, we replace Eq.\ (\ref{eq:burgervl})
by the more conventional notation
\begin{equation}
\frac{\partial \Phi}{\partial \tau} = \nu \nabla^2 \Phi +
\frac{\lambda}{2} (\nabla \Phi)^2 + V({\bf r}, \tau),
\label{eq:burger}
\end{equation}
where initially $\nu = 2 \lambda = T/2g$, and we have absorbed a
factor of $T$ into $V$ so that $\Delta ({\bf k}) = \Delta/(T^2 k)$.
We generalize to correlated disorder $\Delta({\bf k})= \Delta/(T^2
k^{2\rho})$.  It is instructive to generalize the correlator this way,
because then not only does $\rho=1/2$ describe nearly isotropic
splayed columnar disorder, but $\rho=0$ generates the results for
point disorder as well.  This renormalization group will be based
around a perturbation series in powers of $\lambda$.

We first rescale Eq.\ (\ref{eq:burger}) by a scale factor $b$, with
\begin{eqnarray}
{\bf r} & \rightarrow & b {\bf r}, \\
\tau & \rightarrow & b^z \tau, \\
\Phi & \rightarrow & b^{\chi} \Phi,
\end{eqnarray}
where $z$ and $\chi$ will eventually be chosen to keep various
coupling constants fixed under the renormalization procedure.  Since
$z$ describes the ratio of the scaling in the timelike direction with
that in the spacelike direction, $\zeta = 1/z$ is exactly the
wandering exponent that we are looking for.  Upon inserting these
transformations into Eq.\ (\ref{eq:burger}), we see that the equation
remains invariant under these changes provided we rescale $\nu$,
$\lambda$, and $\Delta$ via
\begin{eqnarray}
\label{eq:scalenu}
\nu & \rightarrow & b^{z-2} \nu, \\
\label{eq:scalelambda}
\lambda & \rightarrow & b^{\chi + z -2} \lambda, \\
\label{eq:scaledelta}
\Delta & \rightarrow & b^{2\rho+1-d+z-2\chi} \Delta.
\end{eqnarray}
In the absence of the nonlinearity (\textit{i.e.}, $\lambda = 0$), the
equation becomes completely scale invariant if we choose $z=2$ and
$\chi = (2\rho+3-d)/2$.  However, any small $\lambda$ then rescales
according to
\begin{equation}
\lambda \rightarrow b^{(2\rho+3-d)/2} \lambda,
\end{equation}
and thus the nonlinear term will be relevant for $d<2\rho+3$
(\textit{i.e.},  $d<4$ for splayed columnar disorder and $d<3$ for
point disorder).  For $d>2\rho+3$, we expect the mean field exponents
displayed in Eq.\ (\ref{eq:scalenu}--\ref{eq:scaledelta}) to be
accurate, and $\zeta = 1/2$.  However, in the physical situation of
interest, $d \leq 2\rho+3$ and we expect the scaling exponents to
change due to the nonlinearity.

\subsubsection{Perturbation theory}
\label{sec-pert}

\begin{figure}[tp]
\begin{center}\leavevmode
\includegraphics[width=0.8\linewidth]{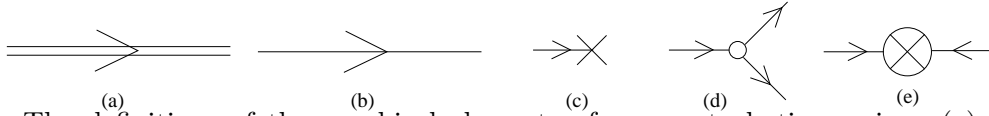}
\caption[The definitions of the graphical elements of our perturbation
series.]{The definitions of the graphical elements of our perturbation
series.  (a) represents the renormalized propagator $G({\bf k},
\omega)$, (b) represents the bare propagator $G_0({\bf k}, \omega)$, (c)
represents the noise $V({\bf k}, \omega)$, (d) represents the
interaction $-\frac{\lambda}{2} \int {\bf q} \cdot ({\bf k} - {\bf
q})$, and (e) represents the disorder correlator $2\Delta({\bf k})$.}
\label{fig:definerg}
\end{center}
\end{figure}

\begin{figure}[tp]
\begin{center}\leavevmode
\includegraphics[width=0.8\linewidth]{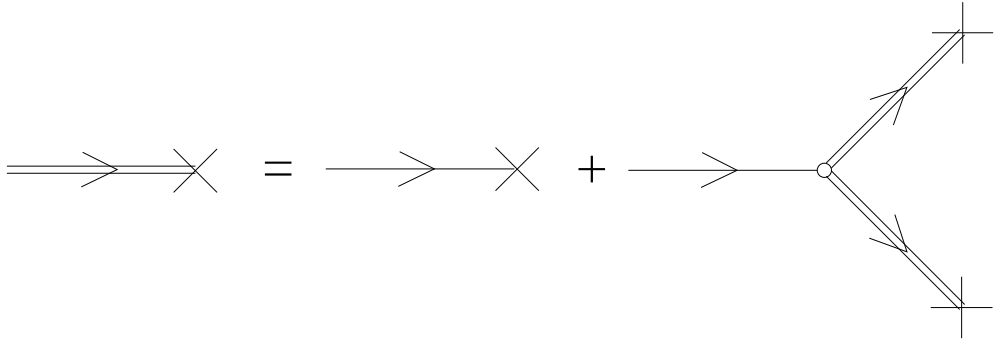}
\caption{Diagrammatic representation of Eq.\ (\ref{eq:burgerft}).}
\label{fig:burger}
\end{center}
\end{figure}

\begin{figure}[tp]
\begin{center}\leavevmode
\includegraphics[width=0.8\linewidth]{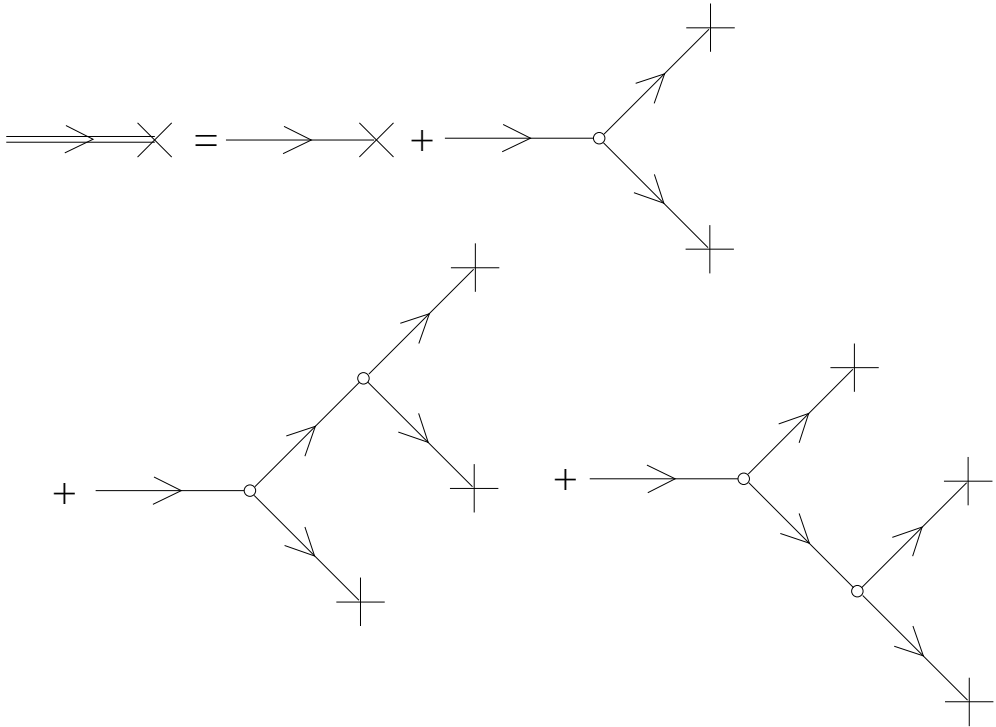}
\caption{Iterated version of Eq.\ (\ref{eq:burgerft}) to second order
in $\lambda$.}
\label{fig:burger2}
\end{center}
\end{figure}

To understand the case $d \leq 2 \rho + 3$, we rewrite Eq.\
(\ref{eq:burger}) in Fourier space via the definition
\begin{equation}
\Phi ({\bf k}, \omega) = \int d^{d-1} {\bf r} \int_0^L d \tau
\Phi({\bf r}, \tau) e^{-i({\bf k} \cdot {\bf r} - \omega \tau)},
\end{equation}
obtaining
\begin{equation}
\Phi ({\bf k}, \omega) = G_0 ({\bf k}, \omega) V({\bf k}, \omega) -
\frac{\lambda}{2} G_0 ({\bf k}, \omega) \int d^{d-1} {\bf k'} \int_0^L
d \omega' {\bf k'} \cdot ({\bf k} - {\bf k'}) \Phi({\bf k'}, \omega')
\Phi({\bf k} - {\bf k'}, \omega - \omega'),
\label{eq:burgerft}
\end{equation}
where
\begin{equation}
G_0 ({\bf k}, \omega) = \frac{1}{- i \omega + \nu k^2}.
\end{equation}
We define a renormalized Green's function $G ({\bf k}, \omega)$ via
\begin{equation}
\Phi ({\bf k}, \omega) \equiv G ({\bf k}, \omega) V({\bf k}, \omega)
\label{eq:fullprop}
\end{equation}
and calculate $G ({\bf k}, \omega)$ perturbatively in $\lambda$.  The
perturbation series can be summarized diagrammatically by giving
graphical representations to the renormalized and bare propagators $G
({\bf k}, \omega)$ and $G_0 ({\bf k}, \omega)$, the disorder $V ({\bf
k}, \omega)$, the interaction, and the disorder correlator $\Delta
({\bf k})$ as shown in Fig.~\ref{fig:definerg}.  As expressed in Eq.\
(\ref{eq:fullprop}), $\Phi ({\bf k}, \omega)$ is represented by a
double arrow followed by a cross to represent the disorder.  Eq.\
(\ref{eq:burgerft}) can then be represented diagrammatically as in
Fig.~\ref{fig:burger}.  We obtain an iterative solution by
substituting Eq.\ (\ref{eq:burgerft}) for each of the $\Phi ({\bf k},
\omega)$ terms appearing on the right hand side of Eq.\
(\ref{eq:burgerft}), thereby obtaining an iterative series in powers
of $\lambda$.  The result for the renormalized propagator is presented
diagrammatically (to second order in $\lambda$) in
Fig.~\ref{fig:burger2}.

\pagebreak

\begin{figure}[tp]
\begin{center}\leavevmode
\includegraphics[width=0.8\linewidth]{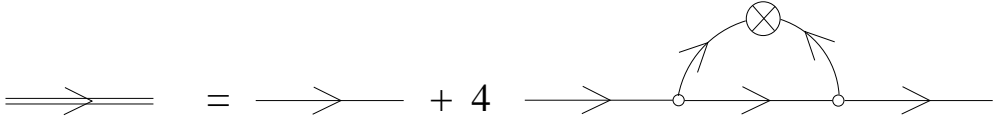}
\caption{Diagrammatic expansion of the renormalized propagator, to the
one-loop order.}
\label{fig:rg}
\end{center}
\end{figure}

We wish to use this perturbation theory to calculate renormalized
versions of the parameters $\nu$, $\lambda$, and $\Delta$ (which we
will denote by $\tilde{\nu}$, $\tilde{\lambda}$, and $\tilde{\Delta}$
respectively).  We define $\tilde{\nu}$ by
\begin{equation}
\lim_{{\bf k} \rightarrow {\bf 0}} G({\bf k}, 0) =
\frac{1}{\tilde{\nu} k^2}
\end{equation}
Upon multiplying Eq.\ (\ref{eq:burgerft}) (or its diagrammatic
equivalent, Fig.~\ref{fig:burger}) by $V(-{\bf k}, - \omega)$ and
averaging over the noise~\cite{log}, we obtain the equation for the
renormalized propagator represented diagrammatically to one loop order
in Fig.~\ref{fig:rg}.

\begin{figure}[tp]
\begin{center}\leavevmode
\includegraphics[width=0.6\linewidth]{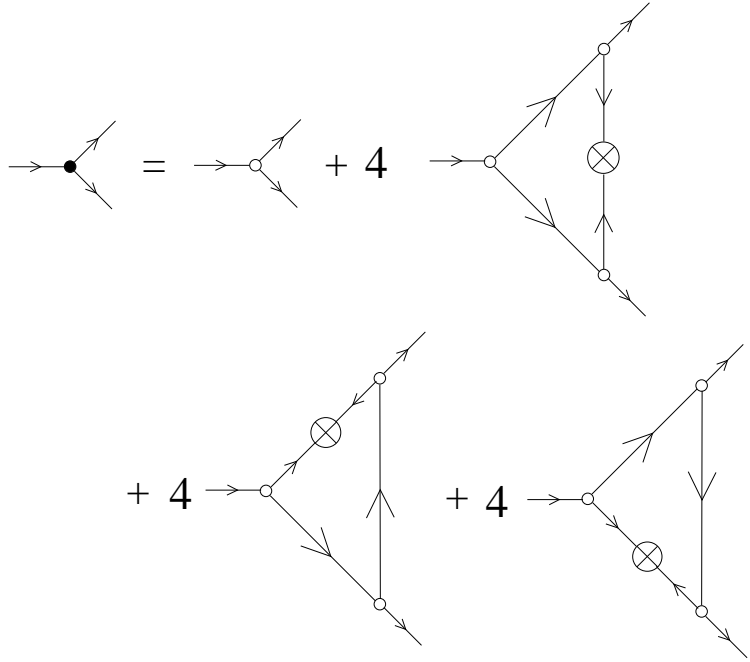}
\caption{Diagrammatic expansion of the renormalized vertex, to the
one-loop order.}
\label{fig:rgb}
\end{center}
\end{figure}

We define $\tilde{\lambda}$ via a renormalized vertex that contains
the effects of the interactions, shown on the left hand side of
Fig.~\ref{fig:rgb}.  The vertex amplitude, in the limit of small ${\bf
k}$, ${\bf q}$, $\omega$, and $\Omega$, is given by
\begin{equation}
- \frac{\tilde{\lambda}}{2} {\bf q} \cdot ({\bf k} - {\bf q}) G_0
({\bf k}, \omega) G_0 ({\bf q}, \Omega) G_0 ({\bf k} - {\bf q}, \omega
- \Omega),
\end{equation}
and serves as the definition of $\tilde{\lambda}$.  Expanding in terms
of the bare quantities, we obtain the equation determining
$\tilde{\lambda}$ to one loop, shown in Fig.~\ref{fig:rgb}.

\begin{figure}[tp]
\begin{center}\leavevmode
\includegraphics[width=0.8\linewidth]{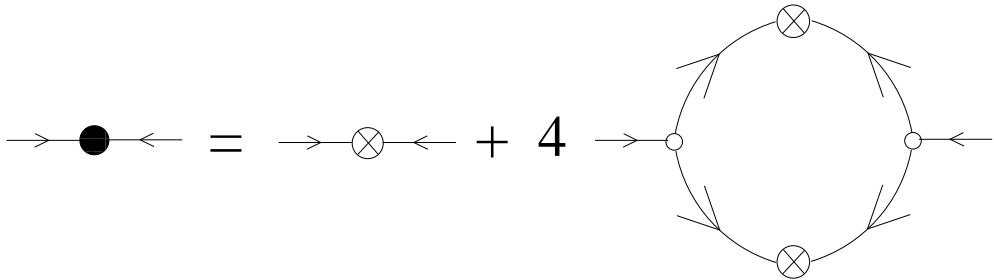}
\caption{Diagrammatic expansion of the renormalized noise correlator,
to the one-loop order.}
\label{fig:rgc}
\end{center}
\end{figure}

We define the renormalized noise correlator $\tilde{\Delta}$ by
\begin{equation}
\overline{\Phi^*({\bf k}, \omega) \Phi({\bf k}, \omega)} \equiv 2
\tilde{\Delta} ({\bf k}) G({\bf k}, \omega) G({-\bf k}, -\omega)
\end{equation}
in the limit of ${\bf k} \rightarrow {\bf 0}$ and $\omega \rightarrow
0$, where
\begin{equation}
\tilde{\Delta} ({\bf k}) = \frac{\tilde{\Delta}}{T^2 k^{2\rho}}.
\end{equation}
Expanding in terms of the bare quantities gives rise to Fig.~\ref{fig:rgc}.

For the case $\rho=0$, diagrams like these are expressed in integral
form and evaluated in detail in Ref.~\cite{FNS} and, \textit{e.g.}, by
Barab\'{a}si and Stanley~\cite{Barabasi}.  The case $\rho \neq 0$ does
not produce any new complications, so we simply report the results:
\begin{eqnarray}
\label{eq:pertnu}
\tilde{\nu} & = & \nu \left[ 1 - \frac{\lambda^2 \Delta}{T^2 \nu^3}
\frac{d-2\rho-3}{4(d-1)} K_{d-1} \int_0^{\Lambda} dq q^{d-2\rho-4}
\right] \\
\label{eq:pertlambda}
\tilde{\lambda} & = & \lambda \\
\label{eq:pertdelta}
\tilde{\Delta} & = & \Delta \left[ 1 + \delta_{\rho,0} \frac{\lambda^2
\Delta}{T^2 \nu^3} \frac{K_{d-1}}{4} \int_0^{\Lambda} dq q^{d-4}
\right],
\end{eqnarray}
where $\Lambda$ is a cutoff in momentum space and $K_d$ is the surface
area of a d-dimensional sphere divided by $(2\pi)^d$.  The nonlinear
coupling $\lambda$ is unrenormalized, as required by Galilean
invariance~\cite{HH+Zhang,FNS}.  Moreover, the noise correlator is
also unrenormalized for any $\rho > 0$; the diagram correcting the
vertex produces only white noise (point disorder).  This can be seen
by noting that the one-loop diagram that renormalizes the noise
correlator (shown in Fig.~\ref{fig:rgc}) is regular as ${\bf k}
\rightarrow {\bf 0}$ because the momenta passing through the disorder
correlator remain finite as ${\bf k} \rightarrow {\bf 0}$.  Since the
other diagrams in Fig.~\ref{fig:rgc} diverge as $1/k^{\rho}$, only
white noise, rather than correlated noise, is produced.  We will
ignore the effect of this white noise for now, since the correlated
noise is more singular, and return to discuss its effects in
Sec.~\ref{sec-white}.

\subsubsection{Renormalization group recursion relations}

All corrections from first order perturbation theory in Eqs.\
(\ref{eq:pertnu})---(\ref{eq:pertdelta}) are well behaved for
$d>2\rho+3$, as expected from the earlier scaling argument.  In lower
dimensions, the renormalization group procedure resums this divergent
perturbation series by integrating over modes with high momentum
$\Lambda e^{-l} < k \leq \Lambda$ and rescaling the resulting
equations by ${\bf k} \rightarrow e^{-l} {\bf k}$.  Upon combining the
scale transformations Eqs.\ (\ref{eq:scalenu})---(\ref{eq:scaledelta})
with the diagrammatic results above we can easily obtain the flow
equations:
\begin{eqnarray}
\label{eq:flownu}
\frac{d\nu}{dl} & = & \nu \left[ z-2 - \frac{\lambda^2 \Delta}{T^2
\nu^3} \frac{d-2\rho-3}{4(d-1)} K_{d-1} \right] \\
\label{eq:flowlambda}
\frac{d\lambda}{dl} & = & \lambda [\chi + z - 2] \\
\label{eq:flowdelta}
\frac{d\Delta}{dl} & = & \Delta \left[ z-2\chi-d+1+2\rho +
\delta_{\rho,0} \frac{\lambda^2 \Delta}{T^2 \nu^3} \frac{K_{d-1}}{4}
\right].
\end{eqnarray}
We can express these in a single flow equation for the combination
$g^2 = \frac{\lambda^2 \Delta}{T^2 \nu^3}$:
\begin{equation}
\frac{dg}{dl} = \frac{3+2\rho-d}{2} g + K_{d-1} g^3
\frac{(3+\delta_{\rho,0})(d-1) - 6\rho - 6}{8(d-1)}.
\label{eq:flowg}
\end{equation}

We look for fixed points of Eq.\ (\ref{eq:flowg}).  We first review
the case $\rho = 0$, \textit{i.e.}, point disorder.  In $d=2$, the
equation reads
\begin{equation}
\frac{dg}{dl} = \frac{1}{2} g - \frac{1}{4} K_{1} g^3,
\end{equation}
which has an unstable fixed point (corresponding to no disorder) at
$g=0$ and a stable fixed point at $g = \sqrt{\frac{2}{K_1}}$.  Upon
inserting this fixed point value into Eqs.\
(\ref{eq:flownu})---(\ref{eq:flowdelta}), we see that the
long-wavelength physics is characterized by $z=3/2$ and $\chi = 1/2$,
giving a wandering exponent of $\zeta = z^{-1} = 2/3$, in agreement
with other results~\cite{FNS,HH+Zhang,MHKZ,HH,FTJ}.  In $d=3$, by
contrast, Eq.\ (\ref{eq:flowg}) reads
\begin{equation}
\frac{dg}{dl} = \frac{1}{8} K_2 g^3.
\end{equation}
The fixed point at $g=0$ is still unstable, but any small disorder
flows off to $g = \infty$, or strong coupling, where our
renormalization group is no longer accurate.  Thus determination of
the wandering exponent for point disorder in three dimensions is
beyond the scope of this method~\cite{HH+Zhang,FNS}.

We now turn to the case $\rho > 0$.  The flow equation for $g$ now
reads
\begin{equation}
\frac{dg}{dl} = \frac{2\rho+3-d}{2} g + \frac{3}{8} \frac{d  - (2\rho
+ 3)}{(d-1)} K_{d-1} g^3.
\end{equation}
For $d < 2\rho +3$, the coefficient of the linear term is positive,
while the coefficient of the cubic term is negative, leading to an
unstable fixed point at $g=0$ and a stable one at
$g=\sqrt{\frac{4(d-1)}{3 K_{d-1}}}$.  Eqs.\
(\ref{eq:flownu})---(\ref{eq:flowdelta}) now lead to
$z=(3+d-2\rho)/3$, $\chi = (2\rho+3-d)/3$, yielding a wandering
exponent of
\begin{equation}
\zeta = \frac{3}{3+d-2\rho}
\end{equation}
For splayed columnar disorder, $\rho=1/2$, and so the wandering
exponent is $\zeta = 3/4$ in two dimensions and $\zeta = 3/5$ in three
dimensions~\cite{MHKZ,HH,FTJ}.

\subsection{Discussion}
\label{sec-white}

One issue that has not been addressed is that of white noise.  Recall
from Sec.~\ref{sec-pert} that the diagram that renormalizes the
disorder correlator $\Delta ({\bf k})$ does not produce any correlated
disorder; however, it does produce white noise.  Thus, even if point
disorder were not initially present, it would be produced by the
renormalization group.  Naively, one would expect that correlated
disorder would dominate over white noise for any $\rho > 0$ since
correlated disorder is more singular than white noise at ${\bf k} =
{\bf 0}$.  This is, in fact the case for $\rho$ sufficiently large;
however, for small $\rho$ the white noise will dominate.  Frey
\textit{et al.}.~\cite{FTJ} have shown that below $d=3+2\rho$, the
renormalization group allows two fixed points---one long-ranged
(dominated by correlated disorder), and one short-ranged (without
correlated disorder).  Moreover, they argue that the fixed point with
the larger dynamic exponent $z$ will be stable, and that if the
long-ranged fixed point is the stable one, its dynamic exponent is
characterized by the exact result
\begin{equation}
z_{\mbox{\scriptsize lr}} = \frac{3+d-2\rho}{3},
\end{equation}
in agreement with the simpler calculation of Sec.~\ref{sec-rgsplay}.
In the case $d=2$, we can use the known result for point disorder
$z=3/2$ to show that the short-ranged fixed point is stable for
$\rho<1/4$, while the long-ranged one is stable for $\rho>1/4$.  This
establishes that for the case of splayed columnar disorder in 2
dimensions, the results are unaffected by the white noise.

The situation is less clear in 3 dimensions because the dynamic
exponent of the short-ranged fixed point, $z_{\mbox{sr}}$, is not
known.  Nevertheless, based on the above considerations, we expect
that there is a curve $\rho_c (d)$ such that for $\rho > \rho_c (d)$,
the long-ranged fixed point is stable, while for $\rho < \rho_c (d)$,
the short-ranged fixed point is stable.  Frey \textit{et al.}.
conjecture that $\rho_c (d) = \frac{d-1}{4}$, based on the fact that
$\rho_c (2) = 1/4$, and $\rho_c (5) = 1$.  (The latter is known from
the fact that in $d=5, \rho = 1$, this equation corresponds to the
Burger's equation with non-conserved noise, previously studied by
Forster \textit{et al}.~\cite{FNS}) This conjecture leads immediately
to a result for $z_{\mbox{\scriptsize sr}}$, namely,
$z_{\mbox{\scriptsize sr}} = \frac{7+d}{6}$, which is in agreement
with the results of Halpin-Healy~\cite{HH} and with numerical
simulations~\cite{Li} in $d=3$.  In three dimensions, this would
therefore imply that splayed columnar disorder is at the boundary
between the regions of stability between the short- and long-ranged
fixed points, and hence that $\zeta = 3/5$ for both point and splayed
columnar disorder.

\section{Dilute vortex lines}
\label{sec-collisions}
\setcounter{equation}{0}

In this section, we show how the $B$ vs.\ $H$ constitutive relation,
which can be measured experimentally~\cite{expt}, follows from a
knowledge of the exponent $\zeta$.  Our central result, that
\begin{equation}
B \sim (H - H_{c1})^{\frac{(d-1) \zeta}{2 (1 - \zeta)}},
\label{eq:central}
\end{equation}
applies whenever the lines are dilute and $0 < \zeta <
1$~\cite{logcorrect}.  However, the prefactor (which is important for
comparison with experiment) will depend somewhat on the experimental
regime.  Here, we first derive the above scaling relation,
generalizing results for point disorder~\cite{expt,Nattermann}, and
then find expressions for the prefactor in various regimes of
temperature and disorder.

Recall that parallel columnar defects localize the vortices, yielding
$\zeta = 0$.  Because the vortices are localized, Eq.\
(\ref{eq:central}) does not apply to this case.  In
Sec.~\ref{sec-columns}, we analyze this case with the boson mapping,
finding
\begin{equation}
B \sim \exp \left[ -\frac{C_d}{(H-H_{c1})^{(d-1)/2}} \right].
\end{equation}

\subsection{$B$ vs.\ $H$ scaling relation with point or splayed
columnar disorder}
\label{sec-bhscaling}

We first review the scaling properties of the free energy.  If we
rescale the system by scaling the $\tau$ direction by a factor $l$,
then the transverse directions rescale by a factor $l^{\zeta}$.
According to Eq.\ (\ref{eq:dprm}), the elastic term of the free energy
then scales by a factor $l^{2 \zeta -1}$.  Because the physics at low
temperatures reflects a balance between the pinning and elastic
energies, we expect that the pinning term of the free energy scales
the same way.  Thus, the pinning energy on a scale $l$ is given by
$U_p (l) \sim l^{2 \zeta - 1}$.

The wandering exponent $\Delta r (l) \sim l^{\zeta}$ (where we define
$\Delta r (l) = \left\{ \overline{[{\bf r}(z + l) - {\bf r}(z)]^2}
\right\}^{1/2}$) describes the transverse wandering of lines at long
scales.  At sufficiently high temperatures, thermal wandering
describes the physics at shorter length scales.  We can then match the
small-scale results onto the large-scale results at the length scales
at which both should be valid.  The exact short length scale mechanism
will depend on the experimental conditions, so we keep it general for
now.  We assume that there is a distance in the $\tau$ direction
$l_c$, a transverse distance $x_c$, and an energy scale $U_c$ above
which the results from Sec.~\ref{sec-burgers} are valid.  (We will
provide expressions for these parameters in Sec~\ref{sec-matching}.)
Then, for $l > l_c$,
\begin{eqnarray}
\label{eq:transscale}
\Delta x(l) & = & x_c \left( \frac{l}{l_c} \right)^{\zeta}, \\
\label{eq:enscale}
U_p(l) & = & U_c \left( \frac{l}{l_c} \right)^{2 \zeta - 1},
\end{eqnarray}
which have the correct long-distance behavior and match with the
required values at $l_c$.

When a finite concentration of vortices enter the sample, their
wandering becomes limited by intervortex collisions at a length scale
given by $\Delta x(l^*) = a_0$, where $a_0$ is the average spacing
between the vortex lines; i.e., at $l^* = l_c (a_0/x_c)^{1/
\zeta}$.  To find the optimal density of vortex lines, we can balance
the energy gain (per unit length) $g (H-H_{c1})/H_{c1}$ of allowing a
vortex line to penetrate with the pinning energy lost (per unit
length) $U_p(l^*)/l^*$ due to collisions~\cite{fermion}.  This yields
a vortex spacing
\begin{equation}
a_0 = x_c \left[ \frac{g l_c (H-H_{c1})}{U_c H_{c1}}
\right]^{\frac{\zeta}{2 (\zeta - 1)}}.
\end{equation}
In this, and subsequent formulae, we neglect dimensionless constants
of order unity.  The field $B$ which penetrates the superconducting
sample is related to $a_0$ via $B = \phi_0/(a_0^{d-1} W^{3-d})$, where
for the case of vortex lines confined to a plate-like geometry (as in
Figs.~\ref{fig:point2d} and~\ref{fig:splay2d}), $W$ is the width of
the sample in the third dimension.  It follows that
\begin{equation}
B = \frac{\phi_0}{x_c^{d-1} W^{3-d}} \left[ \frac{g l_c
(H-H_{c1})}{U_c H_{c1}} \right]^{\frac{(d-1) \zeta}{2 (1- \zeta)}}.
\label{eq:constituitive}
\end{equation}
For flux lines in two dimensions, Eq.\ (\ref{eq:constituitive})
reduces to
\begin{equation}
B = \left\{
\begin{array}{ll}
\displaystyle
\frac{\phi_0 g l_c}{x_c W U_c} \frac{H-H_{c1}}{H_{c1}} & \mbox{for
point disorder ($\zeta = 2/3$)} \\
\displaystyle
\frac{\phi_0 g^{3/2} l_c^{3/2}}{x_c W U_c^{3/2}} \left(
\frac{H-H_{c1}}{H_{c1}} \right)^{3/2} & \mbox{for splayed columnar
disorder ($\zeta = 3/4$).}
\end{array} \right.
\label{eq:bhplanar}
\end{equation}
while in three dimensions we have
\begin{equation}
B = \frac{\phi_0 g^{3/2} l_c^{3/2}}{x_c^2 U_c^{3/2}} \left(
\frac{H-H_{c1}}{H_{c1}} \right)^{3/2}
\end{equation}
for either point disorder~\cite{Nattermann} or splayed columnar
disorder ($\zeta = 3/5$).

\subsection{Physics at shorter length scales}
\label{sec-matching}

We now estimate the values $l_c$, $x_c$, and $U_c$ that appear in
Sec.~\ref{sec-bhscaling}.  The physics at short length scales, before
the effects of disorder build up, is given by application of the naive
scaling analysis of Sec.~\ref{sec-scaling} to Eq.\ (\ref{eq:burger}),
which gives $z=2$, or $\zeta = 1/2$.  The physics is similar to that
of a random walk, dominated by thermal disorder, as a function of the
time-like paramter $l$: $x^2 = \nu l$ for $l < l_c$, where $\nu =
T/2g$.  We expect the system to cross over to the large-scale behavior
when the corrections to this diffusion term become comparable to the
initial value.  Upon generalizing Eq.\ (\ref{eq:pertnu}) so that we
only integrate out to a length scale $x_c$, we see that the criterion
which determines $x_c$ is simply
\begin{equation}
\frac{\lambda^2 \Delta}{T^2 \nu^3} \int_{x_c^{-1}}^{\xi^{-1}} dq
q^{d-2\rho-4} \approx 1,
\end{equation}
where we have neglected factors of order unity.  This leads to
expressions for the crossover parameters
\begin{eqnarray}
\label{eq:xcross}
x_c & = & \left( \frac{T^3}{g \Delta} \right)^{\frac{1}{3+2\rho-d}} \\
\label{eq:lcross}
l_c & = & \frac{g}{T} \left( \frac{T^3}{g \Delta}
\right)^{\frac{2}{3+2\rho-d}} \\
\label{eq:ucross}
U_c & = & T
\end{eqnarray}
provided $d < 3+2\rho$.  The last equality results from noting that
$U$ is unrenormalized when $\zeta = 1/2$.  The inequality $d < 3
+2\rho$ is satisfied for splayed columnar disorder in two or three
dimensions, and for point disorder in two dimensions, but not for
point disorder in three dimensions.  In the latter case, one
finds~\cite{N+LeDoussal}
\begin{eqnarray}
\label{eq:xcross2}
x_c & = & \xi e^{\frac{2\pi T^3}{g \Delta}}, \\
\label{eq:lcross2}
l_c & = & \frac{g \xi^2}{T} e^{\frac{4\pi T^3}{g \Delta}}, \\
\label{eq:ucross2}
U_c & = & T.
\end{eqnarray}

The above results apply at sufficiently high temperatures.  However,
as is evident from Eq.\ (\ref{eq:xcross}), $x_c$ decreases with
decreasing temperature.  If $x_c < \xi$, where $\xi$ is the
(transverse) cutoff provided by the superconducting coherence length,
then the thermal regime is absent entirely.  From Eq.\
(\ref{eq:xcross}), we see that this breakdown occurs for temperatures
$T < T^*$, where
\begin{equation}
T^* = \left(g \Delta \xi^{3+2\rho-d} \right)^{1/3}.
\end{equation}

Below this temperature, we must use a zero-temperature treatment to
determine the characteristic scales $x_c$, $l_c$, and $U_c$.  Consider
the free energy contributions displayed in Eq.\ (\ref{eq:dprm}), given
that the vortex line has typically wandered a transverse distance $x_c
= \xi$ in a longitudinal distance $l_c$.  We assume for simplicity
that $\xi \ll l_c$.  The energy cost of this wandering arising from
the first term of Eq.\ (\ref{eq:dprm}) is approximately $g \xi^2 /
l_c$.  This energy is offset by the line's ability to find a more
hospitable pinning environment.  Let $V(l_c) = \int_0^{l_c} V[{\bf r}
(\tau), \tau] d\tau$ describe the pinning energy of the wandering
line.  The gain in energy due to wandering should be of order the
standard deviation of this zero mean random variable, namely
$\sqrt{\overline{V^2(l_c)}}$.  We have
\begin{eqnarray}
\overline{V^2(l_c)} & = & \int_0^{l_c} d\tau \int_0^{l_c} d\tau'
\overline{V[{\bf r} (\tau), \tau] V[{\bf r} (\tau'), \tau']} \\
& = & l_c \Delta \int \frac{d^{d-1}{\bf k}}{(2\pi)^{d-1}}
\frac{1}{k^{2\rho}}.
\label{eq:naivepert}
\end{eqnarray}
The final integral has an infrared cutoff given by $a_0^{-1}$ and an
ultraviolet cutoff given by $\xi^{-1}$: due to the finite size of the
vortex core, the vortex line only sees a different disorder profile
when it wanders a distance $\xi$.  For $d > 1+ 2\rho$, as is the case
for point disorder in two or three dimensions and for splayed columnar
disorder in three dimensions, the ultraviolet cutoff dominates and we
have
\begin{equation}
\overline{V^2(l_c)} \approx \frac{l_c \Delta}{\xi^{d-1-2\rho}}.
\label{eq:pinning}
\end{equation}
For splayed columnar disorder in two dimensions, we find a logarithmic
correction,
\begin{equation}
\overline{V^2(l_c)} = l_c \Delta \ln (a_0 / \xi).
\label{eq:pinning2}
\end{equation}

Balancing the energy gain due to disorder with the energy loss due to
wandering leads to a characteristic length $l_c$
\begin{equation}
l_c = \left( \frac{g^2 \xi^{d+3-2\rho}}{\Delta} \right)^{1/3}
\label{eq:lzero}
\end{equation}
for $d > 1 + 2 \rho$, and
\begin{equation}
l_c = \left( \frac{g^2 \xi^{4}}{\Delta \ln (a_0 / \xi)} \right)^{1/3}.
\label{eq:lzero2}
\end{equation}
for $d = 1 + 2 \rho$ (splayed columnar disorder in two dimensions).
The corresponding energy scale is 
\begin{equation}
U_c = \left( g \Delta \xi^{3+2\rho-d} \right)^{1/3},
\label{eq:uzero}
\end{equation}
except for splayed columnar disorder in 2 dimensions, where
\begin{equation}
U_c = \left( g \Delta \xi^2 \ln (a_0 / \xi) \right)^{1/3}
\label{eq:uzero2}
\end{equation}
Up to logarithmic corrections, these results match smoothly onto the
high-temperature formulae of Eqs.\
(\ref{eq:xcross})---(\ref{eq:ucross}) in the region where they both
apply, namely, $T \approx \left( g \Delta \xi^{3+2\rho-d}
\right)^{1/3}$.

The above results require $l_c \gg \xi$, an assumption that breaks
down if $\Delta$ is sufficiently large.  In fact, the results observed
for point disorder in two dimensions by Bolle \textit{et
al}.~\cite{expt} indicate that for this experiment, $l_c < \xi$.  The
relevant estimates in this regime are summarized in
Appendix~\ref{sec-kink}.

Combining the results for the matching parameters from Eqs.\
(\ref{eq:xcross})---(\ref{eq:ucross}) and
(\ref{eq:lzero})---(\ref{eq:uzero2}) with the $B$ vs.\ $H$
constitutive relation of Eq.\ (\ref{eq:constituitive}) leads to
\begin{equation}
B = \left\{
\begin{array}{ll}
\displaystyle
\frac{\phi_0 g T}{W \Delta} \frac{H-H_{c1}}{H_{c1}} & \mbox{for $T \gg
(g \Delta \xi)^{1/3}$} \\
\displaystyle
\frac{\phi_0}{W} \left( \frac{g^4 \xi}{\Delta^2} \right)^{1/3}
\frac{H-H_{c1}}{H_{c1}} & \mbox{for $T \ll (g \Delta \xi)^{1/3}$}
\end{array} \right.
\end{equation}
for point disorder in two dimensions (Fig.~\ref{fig:point2d}),
\begin{equation}
B = \left\{
\begin{array}{ll}
\displaystyle
\frac{\phi_0 g^3 \xi}{T^3} e^{\frac{K T^3}{g \Delta}} \left(
\frac{H-H_{c1}}{H_{c1}} \right)^{3/2} & \mbox{for $T \gg (g
\Delta)^{1/3}$} \\
\displaystyle
\frac{\phi_0 g^2 \xi}{\Delta} \left( \frac{H-H_{c1}}{H_{c1}}
\right)^{3/2} & \mbox{for $T \ll (g \Delta)^{1/3}$}
\end{array} \right.
\end{equation}
for point disorder in three dimensions (Fig.~\ref{fig:point3d}),
\begin{equation}
B = \left\{
\begin{array}{ll}
\displaystyle
\frac{\phi_0 g^2}{W \Delta} \left( \frac{H-H_{c1}}{H_{c1}}
\right)^{3/2} & \mbox{for $T \gg (g \Delta \xi^2)^{1/3}$} \\
\displaystyle
\frac{\phi_0 g^2}{W \Delta \ln \left[ \frac{\Delta}{\xi g^2} \left(
\frac{H_{c1}}{H-H_{c1}} \right)^{3/2} \right]} \left(
\frac{H-H_{c1}}{H_{c1}} \right)^{3/2} & \mbox{for $T \ll (g \Delta
\xi^2)^{1/3}$} \\
\end{array} \right.
\end{equation}
for splayed columnar disorder in two dimensions
(Fig.~\ref{fig:splay2d}), and
\begin{equation}
B = \frac{\phi_0 g^2}{\Delta} \left( \frac{H-H_{c1}}{H_{c1}}
\right)^{3/2}
\end{equation}
for splayed columnar disorder in three dimensions
(Fig.~\ref{fig:splay3d}).

\subsection{Crossover between splayed columnar and point disorder:
finite length columns}
\label{sec-finitescaling}

Splayed columnar disorder arising from fission fragments often
consists of columns with a typical length $l_{\mbox{\scriptsize col}}$
that is much smaller than the sample size $L$ (as seems to be the case
in Refs.~\cite{Civale2,Kruzin,Hardy2}).  We then expect to observe a
crossover from the behavior typical of splayed columnar disorder to
that of point disorder sufficiently close to $H_{c1}$.  On scales $l$
such that $l_{\mbox{\scriptsize col}} \ll l \ll L$, the vortex lines
feel the finite size of the columns, and thus the behavior should be
closer to that described by point disorder.  However, for $l_c \ll l
\ll l_{\mbox{\scriptsize col}}$, the vortex lines behave as if the
columns were infinitely long, and thus the behavior is that of splayed
columnar disorder.  In other words, in two dimensions, we expect the
$B$ vs.\ $H$ constitutive relation to be $B \sim (H-H_{c1})$ at very
weak fields, where the length scale between collisions is above
$l_{\mbox{\scriptsize col}}$, and $B \sim (H-H_{c1})^{3/2}$ at
somewhat stronger fields.  In three dimensions, since the constitutive
relation is the same for both point and splayed columnar disorder,
the crossover will appear only in the amplitude of the power law.

In the regime $l \gg l_{\mbox{\scriptsize col}}$, we expect that the
behavior can be described via the methods of Sec.~\ref{sec-bhscaling},
with splayed columnar disorder playing the role of the small-scale
mechanism alluded to near the beginning of Sec.~\ref{sec-bhscaling}.
Specifically, let $x_x$ and $U_x$ be the transverse length scale and
energy at which the behavior will cross over from splayed columnar to
point disorder.  (These will play the role of $x_c$ and $U_c$
respectively, while $l_{\mbox{\scriptsize col}}$ will play the role of
$l_c$.)  Then, applying Eq.\ (\ref{eq:transscale}), we find
\begin{equation}
\Delta x (l) = \left\{
\begin{array}{ll}
\displaystyle
x_c \left( \frac{l}{l_c} \right)^{3/4}, & l_c \ll l \ll
l_{\mbox{\scriptsize col}} \\
\displaystyle
x_x \left( \frac{l}{l_{\mbox{\scriptsize col}}} \right)^{2/3}, & l \gg
l_{\mbox{\scriptsize col}}.
\end{array} \right.
\end{equation}
Matching these formulae at $l_{\mbox{\scriptsize col}}$, we see that
\begin{equation}
x_x = x_c \left( \frac{l_{\mbox{\scriptsize col}}}{l_c} \right)^{3/4}.
\end{equation}
Similarly, by using Eq.\ (\ref{eq:enscale}) and matching at
$l_{\mbox{\scriptsize col}}$, we obtain
\begin{equation}
U_x = U_c \left( \frac{l_{\mbox{\scriptsize col}}}{l_c} \right)^{1/2}.
\end{equation}
Eq.\ (\ref{eq:bhplanar}) then leads to
\begin{equation}
B = \left\{
\begin{array}{ll}
\displaystyle
\frac{\phi_0 g l_{\mbox{\scriptsize col}}}{x_x W U_x}
\frac{H-H_{c1}}{H_{c1}} & \mbox{for sufficiently weak fields} \\
\displaystyle
\frac{\phi_0 g^{3/2} l_c^{3/2}}{x_c W U_c^{3/2}} \left(
\frac{H-H_{c1}}{H_{c1}} \right)^{3/2} & \mbox{for stronger fields.}
\end{array} \right.
\label{eq:finitesplay}
\end{equation}

We need to find what the field strength will be at crossover.
Crossover occurs when the distance between vortex lines $a_0$ becomes
comparable to $x_x$.  In other words, we cross over to the splayed
columnar disorder result at the field at which a vortex line typically
collides with another vortex line every $l_{\mbox{\scriptsize col}}$
in the $z$ direction.  This yields
\begin{equation}
B = \frac{\phi_0}{x_x W} = \frac{\phi_0}{x_c W} \left(
\frac{l_c}{l_{\mbox{\scriptsize col}}} \right)^{3/4}.
\end{equation}
To find the $H$ at which this occurs, we use Eq.\
(\ref{eq:finitesplay}).  Whichever expression we use, the same result
is obtained, which demonstrates the self-consistency of our result,
\begin{equation}
\frac{H-H_{c1}}{H_{c1}} = \frac{U_c}{g \sqrt{l_c l_{\mbox{\scriptsize
col}}}}.
\end{equation}
In summary, we conclude that
\begin{equation}
B = \left\{
\begin{array}{ll}
\displaystyle
\frac{\phi_0 g l_c}{x_c W U_c} \left( \frac{l_c}{l_{\mbox{\scriptsize
col}}} \right)^{1/4} \frac{H-H_{c1}}{H_{c1}}, &
\displaystyle
\frac{H-H_{c1}}{H_{c1}} \ll \frac{U_c}{g \sqrt{l_c
l_{\mbox{\scriptsize col}}}} \\
\displaystyle
\frac{\phi_0 g^{3/2} l_c^{3/2}}{x_c W U_c^{3/2}} \left(
\frac{H-H_{c1}}{H_{c1}} \right)^{3/2}, & 
\displaystyle
\frac{H-H_{c1}}{H_{c1}} \gg \frac{U_c}{g \sqrt{l_c
l_{\mbox{\scriptsize col}}}}.
\end{array} \right.
\end{equation}

The parameters $x_c$, $l_c$, and $U_c$ appearing in these equations
are those of two dimensional splayed columnar disorder, which
dominates at short length scales.  This yields for high temperatures
($T \gg [g \Delta \xi^2]^{1/3}$)
\begin{equation}
B = \left\{
\begin{array}{ll}
\displaystyle
\frac{\phi_0 g^{3/2}}{W l_{\mbox{\scriptsize col}}^{1/4} \Delta^{3/4}}
\frac{H-H_{c1}}{H_{c1}}, &
\displaystyle
\frac{H-H_{c1}}{H_{c1}} \ll \frac{\Delta^{1/2}}{g l_{\mbox{\scriptsize
col}}^{1/2}} \\
\displaystyle
\frac{\phi_0 g^2}{W \Delta} \left( \frac{H-H_{c1}}{H_{c1}} \right)^{3/2}, & 
\displaystyle
\frac{H-H_{c1}}{H_{c1}} \gg \frac{\Delta^{1/2}}{g l_{\mbox{\scriptsize
col}}^{1/2}}
\end{array} \right.
\end{equation}
and for low temperatures ($T \ll [g \Delta \xi^2]^{1/3}$)
\begin{equation}
B = \frac{\phi_0 g^{3/2}}{W l_{\mbox{\scriptsize col}}^{1/4}
\Delta^{3/4} \left[ \ln \left( \frac{\Delta^{3/4} l_{\mbox{\scriptsize
col}}^{1/4}}{g^{3/2} \xi} \frac{H_{c1}}{H-H_{c1}} \right)
\right]^{3/4}} \frac{H-H_{c1}}{H_{c1}} 
\end{equation}
for
\begin{equation}
\frac{H-H_{c1}}{H_{c1}} \ll \frac{\Delta^{1/2} \left\{ \ln \left[
\frac{\Delta}{\xi g^2} \left( \frac{H_{c1}}{H-H_{c1}} \right)^{3/2}
\right] \right\}^{1/2}}{g l_{\mbox{\scriptsize col}}^{1/2}},
\end{equation}
while
\begin{equation}
B = \frac{\phi_0 g^2}{W \Delta \ln \left[ \frac{\Delta}{\xi g^2}
\left( \frac{H_{c1}}{H-H_{c1}} \right)^{3/2} \right]} \left(
\frac{H-H_{c1}}{H_{c1}} \right)^{3/2}
\end{equation}
for
\begin{equation}
\frac{H-H_{c1}}{H_{c1}} \gg \frac{\Delta^{1/2} \left\{ \ln \left[
\frac{\Delta}{\xi g^2} \left( \frac{H_{c1}}{H-H_{c1}} \right)^{3/2}
\right] \right\}^{1/2}}{g l_{\mbox{\scriptsize col}}^{1/2}}.
\end{equation}

\begin{minipage}[t]{6in}
\vspace{0.1in}
\begin{tabular}{|c|c|} 
\hline Vortex lines & Bosons \\ \hline \hline
$g$ &  $m$ \\ \hline
$k_B T$ & $\hbar$ \\ \hline
$L_z$ & $\beta \hbar$ \\ \hline
$(H - H_{c1}) \phi_0 / 4 \pi$ & $\mu$ \\ \hline
$B / \phi_0$ & $n$ (boson density) \\ \hline
\hspace{.3in}Vortex lines in three-dimensional samples \hspace{.3in} &
\hspace{.3in}Two-dimensional bosons \hspace{.3in} \\ \hline
Vortex lines in two-dimensional samples & One-dimensional bosons \\
\hline
Parallel columnar disorder & Point disorder \\ \hline 
\end{tabular}
\vspace{0.1in}
\begin{small}

TABLE I.
Detailed correspondence of the parameters of the vortex line system
with the parameters of the boson system.
\end{small}
\vspace{0.2in}
\end{minipage}

\subsection{$B(H)$ constitutive relation with parallel columnar
disorder}
\label{sec-columns}

The boson mapping~\cite{N88,N+Vinokur,N+LeDoussal,Lehrer+N} is
particularly useful to study vortex lines in the presence of parallel
columnar defects.  The dimensionality of the fictitious bosons is one
lower than that of the superconducting sample, \textit{i.e.}, vortices
in three-dimensional superconductors are described by two-dimensional
bosons, while those in two-dimensional superconductors correspond to
one-dimensional bosons.  Parallel columnar disorder plays the role of
point disorder, while the temperature $T$ plays the role of Planck's
constant $\hbar$, the bending energy $g$ plays the role of the boson
mass $m$, and the sample length $L$ plays the role of $\beta \hbar$
for the bosons.  (See Table I for a summary.)  Since all eigenstates
are localized in one and two dimensional quantum mechanics, $\zeta =
0$ for vortex lines in the presence of disordered parallel columnar
pins in both two and three dimensions.  Thus, Eq.\ (\ref{eq:central})
does not apply: the physics leading to the $B(H)$ constitutive
relation is very different.  Rather than restricting vortex wandering,
intervortex repulsion will assign an energy cost to two vortex lines
that occupy the same localized region.  For $B \ll H_{c1}$, this
energy cost will be prohibitive, and we approximate the effects of the
interaction as prohibiting multiple occupancy of the same
state~\cite{N+Vinokur,Hatano+N}.  Thus, the vortex interactions play
the role of the Pauli exclusion principle, and, in this approximation,
the behavior is the same as for spinless non-interacting fermions.

From Table I, the $n(\mu)$ relationship at $T=0$ for the fictitious
bosons yields the $B(H)$ relationship in the thermodynamic limit $L
\rightarrow \infty$.  But the $n(\mu)$ relationship is simply given by
\begin{equation}
n(\mu) = \int_{-\infty}^{\mu} g(E) dE,
\end{equation}
where $g(E)$ is the non-interacting density of states per unit energy
per unit area.  Thus the $B(H)$ relation in the dilute limit is
determined by the low energy tail of the density of states.

Larkin and Vinokur~\cite{Larkin+Vinokur} determined $B(H)$ in the
following fashion: they assumed a Gaussian disorder potential,
\begin{equation}
\langle V({\bf r}) V({\bf r}') \rangle = \Delta_1 \delta^2 ({\bf r} -
{\bf r}'),
\end{equation}
from which it follows that at low energies~\cite{gausstail},
\begin{equation}
g(E) \sim e^{-2.9 E/E_0},
\label{eq:gaussdos}
\end{equation}
with $E_0 = \Delta_1 g / T^2$ in the vortex line language.  The end
result is that
\begin{equation}
B \sim e^{{\mathcal{N}} \phi_0 (H-H_{c1}) / E_0},
\label{eq:gaussconst}
\end{equation}
where $\mathcal{N}$ is a numerical factor.  However, we do not believe
this to be an accurate description of real flux lines.  The disorder
is taken to be Gaussian, and as such is not bounded below.  Therefore,
there are states at arbitrarily low energy ($E \rightarrow - \infty$),
as Eq.\ (\ref{eq:gaussdos}) shows, leading to vortex penetration at
arbitrarily small fields.  Indeed, according to Eq.\
(\ref{eq:gaussconst}), we would expect a small density of vortex lines
parallel to the $z$-axis to penetrate the sample in the limit $H_z=0$,
and even for $H_z<0$!  This unphysical behavior is an artifact of
choosing a disorder potential that is not bounded below.  To fix this
problem, we choose the pinning potential $V(r)$ from a uniform
distribution over the range $E_0 < V < E_1$.  While the real
distribution of the disorder will be bounded, it may not be of this
form.  Therefore, at the end of this section, we discuss how our
results would be altered by choosing a different (bounded)
distribution.

Clearly, $g(E)$ is bounded from below by $E=E_0$, yielding $H_{c1} =
\frac{4\pi E_0}{\phi_0}$.  The form of the density of states as $E
\rightarrow E_0$ from above is determined by the frequency of large,
rare regions where the disorder potential is always near the bottom of
the band~\cite{Friedberg,N+Shnerb}.  To find $g(E)$, we (see,
\textit{e.g.}, Ref.~\cite{N+Shnerb}) estimate the probability
$p(R,\delta)$ of finding a sphere of radius $R$ with all energies
within $\delta$ of the bottom of the band as
\begin{equation}
p(R,\delta) \approx \left( \frac{\delta}{E_1 - E_0}
\right)^{(R/l_0)^{d_B}} = \exp \left[ \left( \frac{R}{l_0}
\right)^{d-1} \ln \left( \frac{\delta}{E_1 - E_0} \right) \right],
\end{equation}
where $d_B$ is the dimension of the fictitious bosons ($d_B = d-1$)
and $l_0$ is the (microscopic) transverse distance over which the
disorder potential is correlated, \textit{i.e.}, the radius of the
columnar defects.

Because, in the boson mapping, the kinetic energy takes the form (see
Table I) $-\frac{T^2}{2g} \nabla^2$, the low-energy eigenstate
produced by a such an anomalous region will be given approximately by
\begin{equation}
E \approx E_0 + \delta + \frac{cT^2}{gR^2},
\label{eq:rareenergy}
\end{equation}
where $c$ is a numerical factor of order unity.  Therefore, the
probability of finding a state between energy $E$ and $E + dE$ using a
sphere of radius $R$ is given by
\begin{equation}
p(R,E) \sim \left. \frac{\partial p(R,\delta)}{\partial \delta}
\right|_{\delta = E - E_0 - \frac{cT^2}{gR^2}},
\end{equation}
yielding
\begin{equation}
p(R,E) \sim \exp \left[ \left( \frac{R}{l_0} \right)^{d-1} \ln \left(
\frac{E - E_0 - \frac{cT^2}{gR^2}}{E_1 - E_0} \right) \right].
\label{eq:rareprob}
\end{equation}
Note that from Eq.\ (\ref{eq:rareenergy}), since $\delta \geq 0$,the
lower limit $R$ at which it is possible to create a state with energy
$E$ is
\begin{equation}
R = \sqrt{ \frac{cT^2}{g} \frac{1}{E - E_0}}.
\label{eq:smallestrare}
\end{equation}
Upon optimizing Eq.\ (\ref{eq:rareprob}) with respect to $R$, we find
that (up to logarithmic corrections) the maximum occurs at the lower
limit of $R$ given by Eq.\ (\ref{eq:smallestrare}).  Thus, the form of
the density of states at low energies is given by
\begin{equation}
g(E) \sim \exp \left[ - \left( \frac{c' T^2}{g l_0^2} \frac{1}{E - E_0}
\right)^{(d-1)/2} \right].
\label{eq:uniformdos}
\end{equation}

The logarithmic corrections alluded to above will change the factor of
order unity in the exponential, and may introduce pre-exponential
terms.  We do not, however, calculate these effects, as the results
are not independent of the details of the distribution from which the
disorder has been drawn.  In particular, if the distribution is
bounded below and does not vanish too quickly at the lower bound,
only the numerical coefficient of order unity $c'$ and pre-exponential
terms will change; the form of density of states will be the same.
However, if the distribution falls off faster than any power law at
its lower bound (\textit{e.g.}, where the probability of obtaining an
energy $V \sim \exp [-K / (V - E_0)]$), then the exponent $(d-1)/2$ in
Eq.\ (\ref{eq:uniformdos}) may change as well.  Thus, presuming the
disorder potential does not fall off
\textit{too} fast near the bottom of the band,
\begin{equation}
B(H) \sim \exp \left[ - C_d \left( \frac{T}{g l_0} \right)^{d-1}
\left( \frac{H_{c1}}{H-H_{c1}} \right)^{(d-1)/2} \right],
\end{equation}
where $C_d$ is a constant of order unity.

\section{Outlook}
\setcounter{equation}{0}

We expect that the results described above will be valid when the
lines are dilute, \textit{i.e.}, when
$B\,$\raisebox{-.5ex}{$\stackrel{<}{\sim}$}$\,H_{c1}$.  Here we
discuss the outlook for an understanding of vortex lines in disordered
superconductors when
$B\,$\raisebox{-.5ex}{$\stackrel{>}{\sim}$}$\,H_{c1}$, in which case
the effects of interactions between the vortex lines must be taken
into account more carefully.

At high temperatures, generalizations of ``hydrodynamic'' models
described by Marchetti and Nelson~\cite{Marchetti+N} and by Nelson and
Le Doussal~\cite{N+LeDoussal} should describe the lines quite
effectively in the presence of point, parallel columnar, and splayed
columnar disorder.  These models predict a liquid-like state at high
temperatures.  In the presence of point and parallel columnar
disorder, there may be phase transitions to glassy vortex states at
low temperatures both in 2 and in 3
dimensions~\cite{Fisher,Hwa+N+V,Hwa+F,N+Vinokur}.  In the presence of
point disorder in 3 dimensions, two types of glassy phases are
possible.  For sufficiently weak disorder, the vortex lines form a
``Bragg glass'' in which dislocations do not
proliferate~\cite{Carpentier,Kierfeld,Fisher2,Khaykovich,Deligiannis,Fuchs}.
At stronger disorder, dislocations enter the sample.  However, it is
not yet clear if there is a sharp phase transition separating this
``glassy'' state with dislocations from a high-temperature flux
liquid.

The case of splayed columnar disorder in 2 dimensions with dense lines
has been investigated by Devereaux, Scalettar, Zimanyi, and
Moon~\cite{Devereaux}.  They conclude that there is a transition to a
``splay glass'' phase at low temperatures.  The situation is less
clear in three dimensions.  We expect that, in contrast to the case of
point disorder, there will be no Bragg glass in the presence of
splayed columnar disorder: the pinning produced by columns in random
directions attracting the vortex lines is likely to have a much
stronger entangling effect on the vortices than point disorder.  Since
the dislocation-free Bragg glass observed in the presence of point
disorder is only marginally stable to
dislocations~\cite{Carpentier,Kierfeld,Fisher2}, we expect the
analogous system with splayed columnar disorder to be unstable to
dislocations (especially screw dislocations, which cause
entanglement).

\acknowledgements

We are grateful to C. Bolle, D. J. Bishop, G. Blatter, E. Frey,
B. I. Halperin, and T. Hwa for helpful conversations.  One of us (drn)
acknowledges numerous enlightening discussions on the effects of splay
with T. Hwa, P. LeDoussal, and V. Vinokur as part of the collaboration
which led to Ref.~\cite{Hwasplay}.  This research was supported by the
National Science Foundation through Grant No. DMR97-14725 and through
the Harvard Materials Research Science and Engineering Laboratory
via Grant No. DMR98-09363.

\appendix
\renewcommand{\theequation}{\Alph{section}.\arabic{equation}}

\section{Zero-temperature kink regime}
\label{sec-kink}

In this Appendix, we adapt the results of Sec.~\ref{sec-matching} to
the case where the effective temperature is low, so a zero-temperature
approach is applicable, but the effective disorder strength is much
stronger than the elastic energy tending to produce straight vortex
lines.  This appears to be the regime in which the experiments
of~\cite{expt} are performed.  This regime is characterized by the
inequality $l_c < \xi$.  It follows that the free energy of Eq.\
(\ref{eq:dprm}) no longer applies, because it relies on an expansion
of the line tension
\begin{equation}
\sqrt{1 + \left( \frac{d {\bf r}}{d\tau} \right)^2} \approx 1 + \frac{1}{2}
\left( \frac{d {\bf r}}{d\tau} \right)^2.
\end{equation}
In this case, however, we can approximate the wandering on short
scales as being by nearly transverse kinks of distance $\xi$, with an
energy cost of $g \xi$ rather than $g l_c^2 / \xi$.  Balancing this
against the energy gain of pinning of Eqs.\ (\ref{eq:pinning}) and
(\ref{eq:pinning2}), we obtain
\begin{equation}
l_c = \frac{g^2 \xi^{d+1-2\rho}}{\Delta}
\end{equation}
for point disorder in two or three dimensions, and for splayed
columnar disorder in three dimensions, while for splayed columnar
disorder in two dimensions we obtain
\begin{equation}
l_c = \frac{g^2 \xi^2}{\Delta \log (a_0 / \xi)}.
\end{equation}
In either case, we find
\begin{equation}
U_c = g \xi,
\end{equation}
in agreement with the results of Ref.~\cite{expt}.


\begin{references}
\bibitem{nearlyisotropic} In general, the dominant frequency $\omega$
associated with wavevector ${\bf k}$ is $\omega \sim k^{1/\zeta}$,
where $\zeta$ is the wandering exponent discussed below.  We expect
that the replacement $\frac{\Delta}{(k^2 + \omega^2)^{1/2}}
\rightarrow \frac{\Delta}{k}$ is possible whenever $\zeta < 1$.
\bibitem{dimensions}  Throughout this paper, we use the convention
that the number of dimensions is appropriate to the superconducting
application.  In the context of interface growth and surface
roughening, the number of dimensions will be one fewer.
\bibitem{log} Note that it is the logarithm of the partition function,
as expected for quenched random disorder, which enters the average.
\bibitem{logcorrect} Our method cannot exlcude logarithmic
corrections to this result at special values of $d$, similar to those
found for thermal fluctuations in Ref.~\cite{N88}.
\end{references}
\end{document}